\documentstyle[12pt,aaspp4]{article}
\begin{document}
\title{The Structure of Cooling Fronts in Accretion Disks}
\author{Ethan T. Vishniac}
\affil{Department of Astronomy, University of Texas, Austin TX
78712; ethan@astro.as.utexas.edu}
\begin{abstract}
Recent work has shown that the speed of the cooling front in
soft X-ray transients may be an important clue in understanding
the nature of accretion disk viscosity.  In a previous paper
(Vishniac and Wheeler 1996) we derived the scaling law for the
cooling front speed.  Here we derive a similarity 
solution for the hot inner part of disks undergoing cooling.
This solution is exact in the limit of a thin disk, power law opacities,
and a minimum hot state column density which is an infinitesimal
fraction of the maximum cold state density.  For a disk of finite 
thickness the largest error is in the ratio of the mass flow across the cooling
front to the mass flow at small radii.  Comparison
to the numerical simulations of Cannizzo et al. (1995) indicates that 
the errors in other parameters do not exceed
$(c_{sF}/r_F\Omega_F)^q$, that is, the ratio of the sound speed 
at the disk midplane to its orbital velocity, evaluated at the cooling
front, to the $q$th power.  Here $q\approx 1/2$. Its precise value
is determined by the relevant hot state opacity law and the 
functional form of the dimensionless viscosity.
\end{abstract}

\section{Introduction}

The most popular model for soft X-ray transients and dwarf novae
is that both are due to fluctuations in the luminosity from
accretion disks surrounding black holes and white dwarfs, respectively,
and that the specific mechanism that starts an outburst is
a thermal instability associated with the ionization of hydrogen
(for a recent review see Cannizzo 1993a or Osaki 1996).
In addition to their intrinsic interest, these systems may provide
valuable clues to the nature of angular momentum transport
in disks, simply by virtue of being non-stationary systems.
In the thermal instability model these disks make the transition 
back to their quiescent states as their outer parts cool and a 
thermal transition front propagates inward.  Mineshige, Yamasaki, and
Ishizaka (1993)
pointed out that this model will produce the observed exponential
decay of soft X-ray luminosity for soft X-ray transients only if the
radius of the hot phase decreases exponentially, i.e. if the
cooling front speed is proportional to the radius of the uncooled
part of the disk.

The speed of the cooling
front affects the evolution of the hot phase of the disk and
its value will depend on the local physics of the disk, including
the dimensionless viscosity $\alpha$.  More specifically,
Mineshige (1987) and Cannizzo, Shafter and Wheeler (1988) argued that
the cooling front speed is approximately
\begin{equation}
\dot r_F\approx -\alpha_F c_{sF}\left({H\over r}\right) \left({r\over\delta r}\right),
\label{eq:first}
\end{equation}
where $\alpha_F$ is the dimensionless viscosity at the onset of
rapid cooling, $H$ is the disk thickness, $\delta r$ is the width of
the cooling region, $c_s$ is the sound speed in the midplane of the
disk, and the subscripts `F' denote values at the
onset of rapid cooling.  Numerical simulations seemed to show that
$\delta r\sim 0.1 r$.

Recent high resolution work by Cannizzo, Chen and Livio (1995, hereafter
CCL) has shown that this estimate for $\delta r$ was a consequence
of insufficient resolution in the simulations.  They found
$\delta r$ was approximately equal to the geometric mean of the
disk height and radius.  Combining this with equation (\ref{eq:first})
they argued that an exponential decline implied a scaling law
for $\alpha$ of the form
\begin{equation}
\alpha\approx 50\left({c_{sF}\over r_F\Omega_F}\right)^{3/2}\propto 
\left({H\over r}\right)^{3/2}.
\end{equation}
More recently Cannizzo (1996) has shown that a similar result can
be recovered from models of the dwarf nova SS Cygni.

In a previous paper (\cite{VW96}, hereafter VW) we showed that
the cooling front speed was actually independent of the structure
of the disk outside the radius where rapid cooling sets in, and
that the cooling front speed scales as
\begin{equation}
\dot r_F\propto-\alpha_F c_{sF}\left({H\over r}\right)^q,
\label{eq:second}
\end{equation}
where $q$ is determined by the functional form of $\alpha$ and
the opacity law for the hot part of the disk.  Assuming that the
hot phase is thermally stable implies that $q$ lies in the interval
$[0,1]$.  For most reasonable opacity laws $q\approx 1/2$.  This
implies that the exponential luminosity decline of dwarf novae
and soft X-ray transients is not just another test of the form
of $\alpha$, but one that specifically depends only on the physics
of the hot state, thereby eliminating uncertainties regarding
opacities in the cold state and the effects of comparing a largely
ionized hot state to a largely neutral cold one.  Finally,
in VW we managed to show that the cooling front speed is surprisingly
insensitive to the value of $\alpha$ right at the cooling front,
but instead is a measure of its behavior over a broad annulus inward
from the cooling front.

While it is important to see that the scaling law reported in CCL can be
recovered from the simple physical argument presented in VW.  
It would be better
still to see that the numerical coefficients found by CCL can
be reproduced by an analytic solution built around this same
argument.  Furthermore, the numerical solution is not precisely
self-similar and by itself the scaling law does not give us
an understanding of when it might fail.  In this paper
we present a self-similar analytic solution of a cooling wave
in a thin accretion disk and estimate the error involved in
applying it to a disk of finite thickness.  Section 2 contains
a derivation of the thin disk similarity solution.  Section 3
contains a summary of our results, a discussion of the error 
involved in applying this solution to realistic disks, and
a discussion of the implications of this work for the physics
of accretion disks.

\section{Analysis}

The usual sign that a system evolving in space and time 
has a stable self-similar solution is that the spatial distribution
of some dynamically significant variable approaches a self-similar
shape in an experiment or in a numerical simulation.  
In the case of a cooling wave in a disk the column density
in the bulk of the hot phase has a simple power law shape with
a constant exponent (CCL).  However, near the cooling front itself,
in the so-called `precursor' region (cf. VW),
the density drops quickly below an extrapolation of this power
law.  In VW we showed that the ratio of the column density at the 
edge of the precursor region, $\Sigma_p$, to the density at the cooling
front itself, $\Sigma_F$, is given by
\begin{equation}
{\Sigma_p\over\Sigma_F}\approx \left({r_F\Omega(r_F)\over c_{sF}}\right)^q.
\end{equation}
The fact that this ratio varies with $r_F$, and therefore with
time, suffices to show that the shape of the column density
distribution is not time-invariant and that any self-similar
solution to this problem can only be an approximation.

On the other hand, since this column density ratio goes to
infinity as the disk becomes infinitely thin, this result suggests
that it may be useful to approximate the column density at $r_F$
as zero, while retaining a finite mass outflow determined by the
actual conditions at the cooling front.  We will see that this
will allow us to recover an approximate similarity solution.
This similarity solution is, in a sense, imaginary for the
region where $\Sigma<\Sigma_F$, but we can estimate the errors
this procedure introduces by examining the dynamical equations
near the cooling front.  To put it another way, in VW we divided
the hot part of the disk into an interior region, in which the
column density is well-approximated by power laws in radius and
time, and a narrow precursor region in which the density drops
precipitously as the cooling front approaches.  Here we approximate
the entire hot state with a similarity solution, with a thin
outer edge which is unrealistic.  However, we can use this
solution to calculate the evolution of the hot state, and 
use our knowledge of the precursor region to understand the
errors in this solution.

One drawback to this approach is that it is not always
appropriate to ignore the effect of the colder parts of the disk
in calculating the cooling front speed and structure.  More precisely,
material ejected across the cooling front will quickly cool and
slow down.  Conservation of mass implies that the column density
of cold material just outside the cooling front will be comparable
to the column density of hot gas at the edge of the precursor
region.  Since there is a maximum column density for the cold state,
this implies that there is an upper limit to $\Sigma_p/\Sigma_F$,
which is a function of radius and which is comparable to 
$\Sigma_{c,max}/\Sigma_{h,min}$ at that radius.  As a cooling
wave approaches this limit its speed will drop and its structure will 
become dependent on the details of the cold state.  We will ignore this
limitation in what follows, but it has to be taken into account
when the solution presented here implies $\Sigma_p/\Sigma_F$ very
large.  We show at the end of this paper that for the systems
considered so far this does not appear to be an important problem.

The basic equations for mass flow in a thin disk are
\begin{equation}
\partial_t\Sigma=-{1\over 2\pi r}\partial_r(\dot M),
\label{eq:contin}
\end{equation}
\begin{equation}
V_r={2\over \Sigma\Omega r^2}\partial_r\left(r^3\alpha \Sigma
{c_s^2\over\Omega}
\partial_r\Omega\right),
\label{eq:torque}
\end{equation}
and
\begin{equation}
\dot M=2\pi r \Sigma V_r,
\label{eq:mdotdef}
\end{equation}
where $\Sigma$ is the gas column density, $V_r$ is the
radial velocity, $c_s$ is the sound speed, $\alpha$ is the
dimensionless viscosity, and $\dot M$ is the mass flow.

The thermal structure of an optically thick disk is
determined
by its opacity source.  When the opacity law has a simple
power law form, the midplane temperature can be written as
\begin{equation}
T=B_1 \Sigma^{a} \alpha^{b} \Omega^{(2/3)c}.
\label{eq:thermal}
\end{equation}
In the hot portions of the disk, for which the opacity can be
approximated by a Kramers law opacity,
$a=c=3/7$ and $b=1/7$.  Rapid cooling sets in for temperatures
below $T_{min}$, where
\begin{equation}
T_{min}\propto \alpha^{-1.1/7} \Omega^{-{3\over70}},
\label{eq:tmin}
\end{equation}
(cf. CCL, equation (4)).
Consequently, $T_{min}$ has a very weak dependence on radius.
Equations (\ref{eq:thermal}) and (\ref{eq:tmin}) also imply
that the hot phase has a minimum column density $\Sigma_{min}=\Sigma_F$.
However, $\Sigma_{min}$ has a strong dependence on radius.

In what follows we will assume that the dimensionless viscosity $\alpha$
has the form
\begin{equation}
\alpha =\alpha_0 \left({h\over r}\right)^n,
\label{eq:alphdef}
\end{equation}
where $n\approx 3/2$ is required to get consistency with the nearly
exponential decay of flux from soft X-ray transients following  
an outburst.  This is not the only form for $\alpha$ that will
produce an exponential luminosity decay.  The luminosity evolution
is primarily sensitive to the variation of $\alpha$ with radius
and the same result can be obtained from $\alpha\propto r^{2/3}$
(\cite{VW96}) but equation (\ref{eq:alphdef}) is
is the only functional form for $\alpha$
which is consistent with all available phenomenological constraints.
A more troublesome question is whether or not $\alpha$ can be written
as a purely local function or whether it depends in some manner
on the global state of the disk.

We are looking for an asymptotic state in which the interior region of
the disk maintains a self-similar mass distribution approximately 
characterized by a mass accretion rate $\dot M_0(t)$ which varies
slowly with time.  We can write this as
\begin{equation}
\dot M(r,t) =\dot M_0(t) F(y),
\label{eq:masst}
\end{equation}
where 
\begin{equation}
y\equiv {r\over r_F}.
\end{equation}
At small radii the disk will be very close to an equilibrium
state with an approximately constant $\dot M=\dot M_0(t)$, 
so $F(0)=1$.  At large radii the
mass flux will be positive, that is outward, towards the cooling front, 
so $F(1)$ will be negative.  Using equations (\ref{eq:thermal}) and
(\ref{eq:alphdef}) we can define a constant $C_0$ such that
\begin{equation}
\alpha \Sigma c_s^2=C_0\Sigma^{1/q} r_F^{-s} y^{-s},
\label{eq:codef}
\end{equation}
where
\begin{equation}
{1\over q}=1+\left(1+{n\over2}\right){a\over1-{n\over 2}b},
\end{equation}
and
\begin{equation}
s=1-\left(1+{n\over2}\right){1-c\over1-{n\over 2}b}.
\end{equation}

Our choice for the form of equation (\ref{eq:masst}) is motivated
by our previous work in VW in which we showed that the 
mass transfer rate everywhere between the
cooling front and the inner edge of the disk will vary only by
a factor of order unity, although it will change sign.  
This is fairly obvious for the precursor region,
since the mass entering this narrow region as the cooling front advances
has to be approximately balanced by the mass ejected across the 
cooling front.  The fact that the asymptotic state of the cooling
front is one in which the front advances slowly enough to allow the 
interior region of the disk to evolve, implies that the velocity of the
cooling front is directly tied to the accretion velocity in the outer
parts of the interior solution.  This in turn implies that the mass
transfer rate inward in the interior solution is proportional to the
mass transfer rate outward through the cooling front.  It follows that
a general description of the cooling disk is most easily obtained by
referring the mass transfer rate at all radii to its value at 
the inner edge.  (The mass transfer rate across the cooling front 
would be an equally good choice.)

Using these definitions in equation (\ref{eq:torque}) we find that
\begin{equation}
\Sigma=\left({-r_F^s \Omega_F\dot M_0\over 6\pi C_0} y^{s-2}\int_0^y F(\tilde y)
{d\tilde y\over \tilde y^{1/2}}\right)^q.
\label{eq:sigint}
\end{equation}
This expression can be substituted into equation (\ref{eq:contin}) to
obtain
\begin{equation}
q\Sigma \left(\partial_t\ln \dot M_0 +\partial_t \ln r_F\left({1\over2}-
{F(y)y^{1/2}\over \int_0^y F(\tilde y)\tilde y^{-1/2}d\tilde y}\right)\right)
={-\dot M_0\over 2\pi r_F^2}y^{-1}\partial_y F.
\end{equation}
It is convenient to rewrite this as
\begin{equation}
-C_2 y\left(y^{s-2}\int_0^y F(\tilde y){d\tilde y\over\tilde y^{1/2}}\right)^q
\left(1+C_1\left(1-{2y^{1/2}F(y)\over\int_0^y F(\tilde y)\tilde y^{-1/2}
d\tilde y}\right)\right)=\partial_yF,
\label{eq:dfint}
\end{equation}
where
\begin{equation}
C_1\equiv {\partial_t \ln r_F\over2\partial_t \ln \dot M_0},
\label{eq:c1}
\end{equation}
and
\begin{equation}
C_2\equiv -\partial_t\ln \dot M_0 \left({2\pi q r_F^2\over -\dot M_0}\right)
\left({-\dot M_0\Omega_F r_F^s\over 6\pi C_0}\right)^q.
\label{eq:c2}
\end{equation}

Our assumption that the hot phase of the disk can be described by a 
self-similar
solution is equivalent to requiring that $C_1$ and $C_2$ be constant, so
that equation (\ref{eq:dfint}) is time-invariant.
In fact, $C_1$ can be determined directly if we know the scaling of
$\dot r_F$ with $r_F$ and use the
requirement that the various time-varying factors in $C_2$ combine to
give a constant product.  We will return to this point later.  
The value of $C_2$ is determined by requiring that the solution to
equation (\ref{eq:dfint}) satisfy the boundary conditions.  
In other words, it is an eigenvalue of equation (\ref{eq:dfint}).

What are the outer boundary conditions for
the similarity solution?  The condition that
the column density vanishes at $r_F$ implies, using equation
(\ref{eq:sigint}), that
\begin{equation}
\int_0^1 F(\tilde y){d\tilde y\over \tilde y^{1/2}}=0.
\label{eq:outer}
\end{equation}
The mass flow at the outer edge of the disk,
$\dot M_F$, is less precisely determined, since it depends
on the gradient of $\Sigma$ at the cooling front. However, given that
the cooling front is defined by the onset of extremely rapid cooling,
it follows that the radial scale for changes in $\Sigma$ will be close
to the disk height and we can write the radial velocity at the cooling
front as
\begin{equation} 
V_F=\Delta \alpha_F c_{sF},
\label{eq:velf}
\end{equation} 
where $\Delta$ is a constant of order unity.  Here we are interested
in the mass flux at an imaginary point where $\Sigma=0$, but we can
approximate it with the value at $\Sigma=\Sigma_F$.
In CCL's simulation they
found $\Delta\approx 1/6$, but the exact number will depend to some extent 
on the uncertain turbulent radial transport of heat near the cooling front.
In any case we have
\begin{equation}
\dot M_F= 2\pi r_F \Sigma_F \Delta\alpha_F c_{sF}.
\end{equation}
This is not a constraint on $F(1)$, which is the ratio between this
mass flux and the undetermined value of $\dot M_0(t)$.
However, we can use this result to rewrite $C_2$.
Starting with the definition in equation (\ref{eq:c2})
we have
\begin{eqnarray}
C_2&\equiv -\partial_t\ln \dot M_0 \left({2\pi q r_F^2\over -\dot M_0}\right)
\left({-\dot M_0\Omega_F r_F^s\over 6\pi C_0}\right)^q\cr
&=-\partial_t \ln r_F {1\over2C_1}\left({2\pi q r_F^2\Sigma_F\over-\dot M_0}\right)
\left({-\dot M_0\Omega_F\over 6\pi \alpha_F\Sigma_Fc_{sF}^2}\right)^q\cr
&={1\over2C_1}\left({2\pi q r_F\dot r_F\Sigma_F\over \dot M_0}\right)
\left({-\Delta r_F\Omega_F\over3c_{sF}F(1)}\right)^q\cr
&={q\over2C_1}\left({\dot r_F\over V_F}\right)F(1)
\left({r_F\Omega_F\over c_{sF}}\right)^q\left({-\Delta\over3F(1)}\right)^q.
\label{eq:c2bet}
\end{eqnarray}
We see from this that
\begin{equation}
\dot r_F={-6C_1C_2\over q}\left({-\Delta\over3F(1)}\right)^{1-q}
\alpha_F c_{sF}\left({c_{sF}\over r_F\Omega_F}\right)^q,
\label{eq:rfdot}
\end{equation}
in agreement with the scaling derived in VW.

In order to evaluate $C_1$ from its definition in equation (\ref{eq:c1}) we
use the fact that in the asymptotic state $\dot M_F$
has a fixed ratio to $\dot M_0$, so that
\begin{equation}
\dot M_0={2\pi r_F\Delta \alpha_F c_{sF}\Sigma_F\over F(1)}.
\end{equation}
Using equation (\ref{eq:thermal}) and our assumed form for $\alpha$ 
we find that
\begin{equation}
\Sigma_F\propto T_F^{(1-b(n/2))/a} r_F^{(c-b(n/2))/a},
\end{equation}
so that
\begin{equation}
\dot M_0\propto r_F^{((1-b(n/2))/q+c-1)/a} T_F^{(1+bn)/2+(1-b(n/2))/a}.
\end{equation}
Substituting this into the definition of $C_1$ we find that
\begin{equation}
C_1={aq\over 2\left((1-b{n\over2})+(c-1)q+\epsilon q(a(1+bn)/2+(1-bn/2))\right)},
\end{equation}
where
\begin{equation}
T_F\propto r_F^\epsilon.
\end{equation}
We can find the value of $\epsilon$ by using equation (\ref{eq:tmin}). 
It is
\begin{equation}
\epsilon={9-11n\over 140+11n},
\end{equation}
which is typically a small number.
We note that if $n=3/2$ and we can assume a Kramers law opacity,
then
$\epsilon=-0.048$ and $C_1=0.2109$.  This value of $n$ is consistent
with the observed approximately exponential luminosity decay, although any 
value near it would be as well.  

If we integrate equation (\ref{eq:dfint}) for the case with $n=3/2$ and
Kramers opacities, and determine $C_2$ by requiring that $f$ satisfy the
boundary condition given by equation (\ref{eq:outer}), then we
find that
\begin{equation}
\dot M_F = -2.88\dot M_0,
\end{equation}
\begin{equation}
\dot r_F = -0.94\alpha_F c_{sF}\left({c_{sF}\over r_F\Omega_F}\right)^q,
\end{equation}
and $\dot M=0$ at $r=0.36 r_F$.  The cooling front velocity goes
as $\Delta^{1-q}$.  Here we have used $\Delta\approx 1/6$ (from
CCL).  CCL obtained $\dot M_F\sim 2 \dot M_0$ and found a zero
in $\dot M$ at $r\approx 0.38 r_F$.  Assuming that the numerical
solutions of CCL are more accurate than the similarity solution,
the errors in the similarity solution are of order 5 to 10\% for
an example in which $(H/r)^q$ is slightly less than $0.1$.  
The only exception is in
the value of $F(1)$, which seems to be too large by as much as $40$\%.
{}For illustration we
show $\Sigma$, $\dot M$, and the surface temperature, $T_s$, 
as a function of $r$ for the similarity solution in figure 1.

\section{Error Evaluation and Conclusions}

We have found a similarity solution that describes the progress
of a cooling wave in an infinitely thin accretion disk whose hot
state is governed by power law opacities.  For a disk of finite
thickness, the nonzero value of the critical column density 
for the onset of rapid cooling introduces an error at the outer
edge of the similarity solution.  The free parameters of this
solution are almost entirely determined by the physics of accretion
and the hot state opacity, the sole exception being the parameter
$\Delta$, which is determined by the structure of the rapid cooling
zone.  In practice, this means that at present 
$\Delta$ is determined from numerical simulations.  
We can see, by comparison with the numerical simulations of CCL, 
that if we take a Kramers opacity and $n=3/2$ we get a
solution which is
a fair description of the progress of a cooling wave in a disk
surrounding a soft X-ray transient.  Looking at our results in figure
1, we can see that the temperature profile of the hot phase is almost
a perfect power law, with a sharp cutoff near the outer edge.  Clearly,
it will be difficult to observe the cutoff structure directly.
However, this model predicts that the power law temperature
profile of the inner region will be well above the critical
temperature for the thermal instability at the cooling front.  This
should be a testable prediction of this model, although this effect
will be more dramatic in the later stages of the luminosity decline.  
Conversely,
a significantly broader temperature transition profile than the one seen 
here would be an indication of nonlocal effects in $\alpha$.  

In order to evaluate
the size of the errors induced by using this solution, 
we need to examine the structure of the cooling wave near the cooling
front.  In this region the state variables will have a radial
scale length much less than $r$, so the dynamical equations
can be approximated by saving only radial derivatives of $\dot M$
and $\Sigma$.

With this in mind we can write
\begin{equation}
\partial_t\Sigma(r)\approx -\dot r_F\partial_r\Sigma.
\end{equation}
Invoking equation (\ref{eq:contin}) we have 
\begin{equation}
-\dot r_F\partial_r\Sigma={-1\over 2\pi r }\partial_r\dot M.
\end{equation}
We can integrate this from $\Sigma=0$ to $\Sigma=\Sigma_F$  to obtain
\begin{equation}
\Delta\dot M\approx -\dot r_F 2\pi r_F\Sigma_F={-\dot r_F\over V_F}\dot M_F,
\label{eq:sigext}
\end{equation}
where $\Delta \dot M$ is the amount by which the mass outflow at the
edge of the similarity solution, where $\Sigma=0$, exceeds the
mass outflow from the actual cooling front, where $\Sigma=\Sigma_F$.
{}For the particular case examined by CCL we have
\begin{equation}
\Delta\dot M\approx 5.4 \left({c_{sF}\over r_F\Omega_F}\right)^q\dot M_F,
\end{equation}
which will be somewhat less than half for $(c_{sF}/r_F\Omega_F)$ between
$10^{-2}$ and $10^{-3}$.

Although this amounts to a large correction to $\dot M$, in agreement
with the comparison between CCL and the prediction of the
similarity solution, it does not imply that the behavior of $\dot M$
with time is subject to a similarly large correction.  
Assuming a Kramers law opacity and $n=1.5$ we can look at the
evolution of the
mass flow through the point where $\Sigma=\Sigma_F$.  We find
that
\begin{eqnarray}
t\partial_t\ln(\dot M_F-\Delta\dot M)&=t\partial_t \ln\left[\dot M_0
\left(1-{0.94\over\Delta}\left({c_{sF}\over r_F\Omega_F}\right)^q\right)\right]\cr
&=t\partial_t\dot M_0-{0.94\over\Delta}t\partial_t\left({c_{sF}\over r\Omega_F}\right)^q\cr
&=t\partial_t \ln\dot M_0\left[1-{1.88C_1\over\Delta}r_F\partial_{r_F}
\left({c_{sF}\over r_F\Omega_F}\right)^q\right]\cr
&=t\partial_t\ln\dot M_0\left(1-0.6\left({c_{sF}\over r_F\Omega_F}\right)^q\right).
\end{eqnarray}
In other words, we get only a small error when we treat the time
evolution of the mass flow through the $\Sigma=\Sigma_F$ radius
as though it were strictly proportional to the mass flow across
the outer surface of the similarity solution.  Since the realistic
condition on the mass flux is that it is fixed at $\Sigma=\Sigma_F$,
rather than at $\Sigma=0$, the negative sign in the above equation
actually implies that the realistic solution will give
$ t\partial_t \ln \dot M_0$ slightly more than the one predicted
from the similarity solution.  For the calculation of CCL the 
discrepancy should be of order a few percent, which is not
significant.

Similarly we can estimate the error we make in equating the
radius of the cooling front with the $\Sigma=0$ surface by
integrating equation (\ref{eq:torque}) using equations
(\ref{eq:mdotdef}) and (\ref{eq:codef}).  We obtain
\begin{equation}
\Delta r = r_F {3\over\Delta} {c_{sF}\over r_F\Omega_F}.
\end{equation}
{}For the calculation of CCL this implies an error of a few
percent.

CCL pointed out that comparing the observed e-folding time
for the soft X-ray luminosity in X-ray transients to the
results of their simulation gives a dimensionless viscosity
of 
\begin{equation}
\alpha\approx 50 \left({h\over r}\right)^{3/2}.
\end{equation}
This coefficient of $50$ is surprisingly large, since one would
expect that almost any theory which gave the correct scaling
law would involve a coefficient of order unity.  
This estimate of the coefficient is 
roughly inversely proportional to the minimum
hot state temperature, which is fixed by the physics of
the opacity in the hot state.  The only other obvious
way to get a different result is by using a different
value of $\Delta$, the ratio of the outflow
velocity at the cooling front to $\alpha_F c_{sF}$.
However, we see from equation (\ref{eq:rfdot}) that this
enters into the expression for the speed of the cooling
front only as $\sim\Delta^{1/2}$.  Even if we replace the
value calculated by CCL, $\Delta\sim 1/6$, with $\Delta=1$
the estimate of the coefficient in the expression for $\alpha$
only drops to $\sim 20$.

One last point we need to consider is under what circumstances
we can ignore the structure of the cold region in discussing the
velocity and structure of the cooling front.  From the equation
of continuity we see that the column density in the cold disk
region, just outside the cooling front, will be of order 
$\sim \Sigma_F (V_F/-\dot r_F)$.  If this exceeds the maximum
stable column density for the cold gas, then the solution
described here is inapplicable, and the cold region will affect the velocity
and structure of the cooling wave.  The value of the maximum 
cold state column density is somewhat uncertain.  Here
we will use the results of Cannizzo (1993b).  Assuming $n\approx 3/2$,
we get
\begin{equation}
{\Sigma_{c,max}\over \Sigma_F}=6.2 \left({\alpha_F\over 0.1}\right)^{-0.31}
r_{10}^{-0.036},
\end{equation}
where $r_{10}$ is the front radius in units of $10^{10}$ cm.
We need this to be larger than $V_F/-\dot r_F$ which is
\begin{equation}
{V_F\over -\dot r_F}\approx \Delta \left({50\over\alpha_F}\right)^{q/n}.
\end{equation}
We see from these expressions that this condition is generally
satisfied, although not by a large factor, with very little
dependence on $\alpha_F$ or $r_{10}$.  Whether or not it will be
satisfied for substantially different models for $\alpha$ or 
disk opacities depends on the specifics of the models.

In sum, the similarity solution described in this paper will give
a good description of the progress of a cooling wave in a hot
disk as long as four conditions are met.  First, the outflow
velocity across the cooling front needs to be substantially
greater than the cooling front velocity.  In practice, this
means that the hot
part of the disk needs to have a height to radius ratio of
order $10^{-2}$ or less for the thermal instability associated
with hydrogen ionization.  If this condition is violated then
the errors in the solution can become large.  Other
thermal instabilities will involve slightly different conditions,
depending on the values of $q$ and $\Delta$.  Second, the
the opacity of the hot disk material, for a few e-foldings
interior to the cooling front, has to be described by a 
power law.  The explicit solution here assumed a Kramers law,
but any power law approximation will yield a similar solution.
Third, the $\alpha$ needs to be purely a function of local
disk parameters.  This last condition is significant, inasmuch
as the only prediction for $\alpha$ which self-consistently
allows for the effects of local magnetic instabilities 
and gives the correct scaling law
is based on an explicitly non-local model (cf. \cite{VJD90},
\cite{VD92}).  We plan an extension of this work to cover
this model and to see if an observably different cooling
front structure can be recovered from it.  Fourth, neglecting
the structure of the disk outside the radius where rapid cooling
sets in is appropriate only when the ratio of $V_F$ to $\dot r_F$
is not so large as to imply an unphysically large column density
in the cold gas.  This does not seem to be an important limitation
for the systems considered to date, but this conclusion is sensitive
to the details of the cold disk structure.

\acknowledgements
I am happy to acknowledge helpful conversations with
John Cannizzo and J. Craig Wheeler.  This work has
been supported in part by NASA through grant
NAG5-2773.

\clearpage
\begin{figure}
\plotone{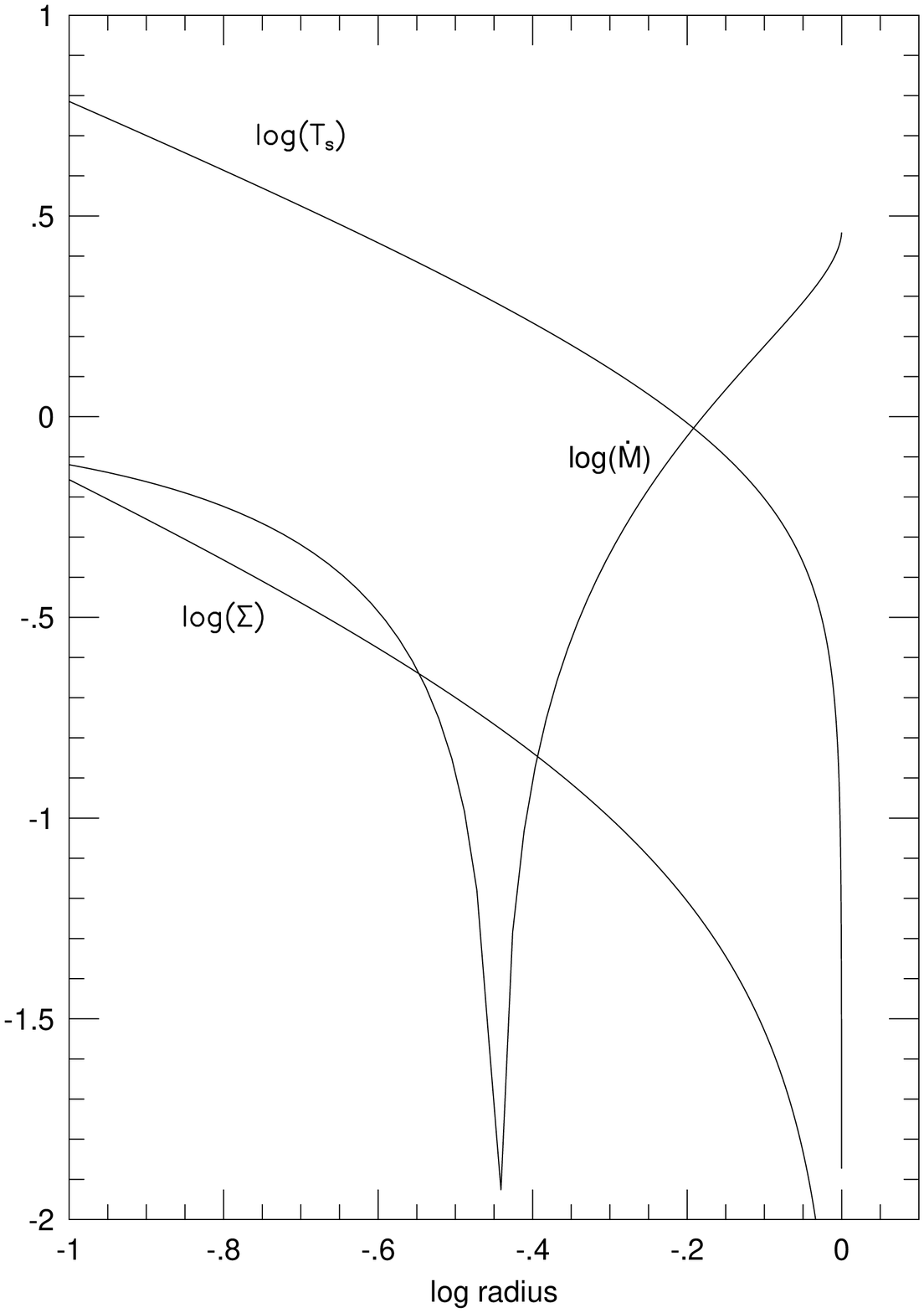}
\caption{Surface temperature, column density, and mass flow as a
function of radius for the cooling wave similarity solution using
Kramers opacity and $n=3/2$.  The vertical axis is in arbitrary
units, but $\dot M$ is normalized to $1$ at small radii.}
\end{figure}
\end{document}